\documentclass[reprint,amsmath,amssymb,aps,pre]{revtex4-1}
\usepackage[utf8]{inputenc}
\usepackage{graphicx}% Include figure files
\usepackage{dcolumn}% Align table columns on decimal point
\usepackage[normalem]{ulem}
\usepackage{mathrsfs,amsmath}
\usepackage{bm}
\usepackage{xcolor}
\usepackage{longtable}
\usepackage{url}
\usepackage{tikz}
\usepackage{natbib}
\usepackage[hidelinks]{hyperref}
\hypersetup{
  colorlinks   = true, %Colours links instead of ugly boxes
  urlcolor     = blue, %Colour for external hyperlinks
  linkcolor    = blue, %Colour of internal links
  citecolor   = teal %Colour of citations
}

\begin{document}
\title{Frequency- and dissipation-dependent entanglement advantage in spin-network Quantum Reservoir Computing}
\author{Youssef Kora}
\thanks{Department of Physics and Astronomy, University of Calgary, Calgary, Alberta, Canada}
\thanks{Institute for Quantum Science and Technology, University of Calgary,  Calgary,  Canada}
\thanks{Hotchkiss Brain Institute, University of Calgary,  Calgary,  Canada}
\author{Hadi Zadeh-Haghighi}
\thanks{Department of Physics and Astronomy, University of Calgary, Calgary, Alberta, Canada}
\thanks{Institute for Quantum Science and Technology, University of Calgary,  Calgary,  Canada}
\thanks{Hotchkiss Brain Institute, University of Calgary,  Calgary,  Canada}
\author{Terrence C Stewart} 
\thanks{National Research Council, University of Waterloo Collaboration Centre, Waterloo, Ontario, Canada}
\author{Khabat Heshami} 
\thanks{Theory and Simulation Group, NRC, Ottawa, Canada}
\thanks{Department of Physics, University of Ottawa, Ottawa, Ontario, Canada}
\thanks{Department of Physics and Astronomy, University of Calgary, Calgary, Alberta, Canada}
\thanks{Institute for Quantum Science and Technology, University of Calgary,  Calgary,  Canada}
\author{Christoph Simon}
\thanks{Department of Physics and Astronomy, University of Calgary, Calgary, Alberta, Canada}
\thanks{Institute for Quantum Science and Technology, University of Calgary,  Calgary,  Canada}
\thanks{Hotchkiss Brain Institute, University of Calgary,  Calgary,  Canada}
\date{\today}

\begin{abstract}
 We study the performance of an Ising spin network for quantum reservoir computing (QRC) in linear and non-linear memory tasks.  We investigate the extent to which quantumness enhances performance by monitoring the behaviour of quantum entanglement, which we quantify by the partial transpose of the density matrix. In the most general case where the effects of dissipation are incorporated, our results indicate that the strength of the entanglement advantage depends on the frequency of the input signal; the benefit of entanglement is greater with more rapidly fluctuating signals, whereas a low-frequency input is better suited to a non-entangled reservoir. This may be understood as a condition for an entanglement advantage to manifest itself: the system's quantum memory must survive for long enough for the temporal structure of the input signal to reveal itself. We also find that quantum entanglement empowers a spin-network quantum reservoir to remember a greater number of temporal features.
\end{abstract}
\maketitle

\section{Introduction}

At the heart of machine learning lie the mechanisms that enable machines to understand, process, and act upon temporal data. The importance of these mechanisms grows rapidly as we push further into the information age, where the complexity and volume of data are scaling at unprecedented rates. Among the principal hurdles in this pursuit is the inherent limitation in processing speeds of conventional computers caused by the segregation of processing and memory units, known as the von Neumann bottleneck  \cite{von_Neumann_1993}. This stands in contrast with biological systems, which handle real-time information processing with exceeding computational efficiency and minimal energy expenditure \cite{Eliasmith_2012,Stewart_2012}. \\ \indent
Within this landscape, the paradigm of reservoir computing (RC) \cite{jaeger_2004,maass_2002,verstraeten_2007} emerges as a powerful mechanism, characterized by a high-dimensional dynamical structure known as a reservoir. This system, upon receiving streams of input, engenders transient dynamics featuring a fading memory capacity and the ability to perform nonlinear processing on input data. The complex internal dynamics of the reservoir render RC exceptionally well-suited for machine learning tasks that demand memory-retention capabilities, such as speech recognition, stock market prediction, and autonomous motor control for robots \cite{Tanaka_2019}. Traditional approaches to RC have been based on either randomly-connected artificial neural networks or through spiking neural networks \cite{Nicola_2017}. Physical implementations of RC have been realized with photonics \cite{Vandoorne_2014,Antonik_2017,Larger_2017,Sunada_2021,Garc_2023}, phonons \cite{Dion_2018,Meffan_2023}, magnons \cite{Papp_2021,Gartside_2022,K_rber_2023}, spintronics \cite{Torrejon_2017,Furuta_2018,Tsunegi_2018}, and nanomaterials structured in neuromorphic chips \cite{Stieg_2011,Yaremkevich_2023}.
\\ \indent
\begin{figure}[h]
\centering
\includegraphics[width=0.5\textwidth]{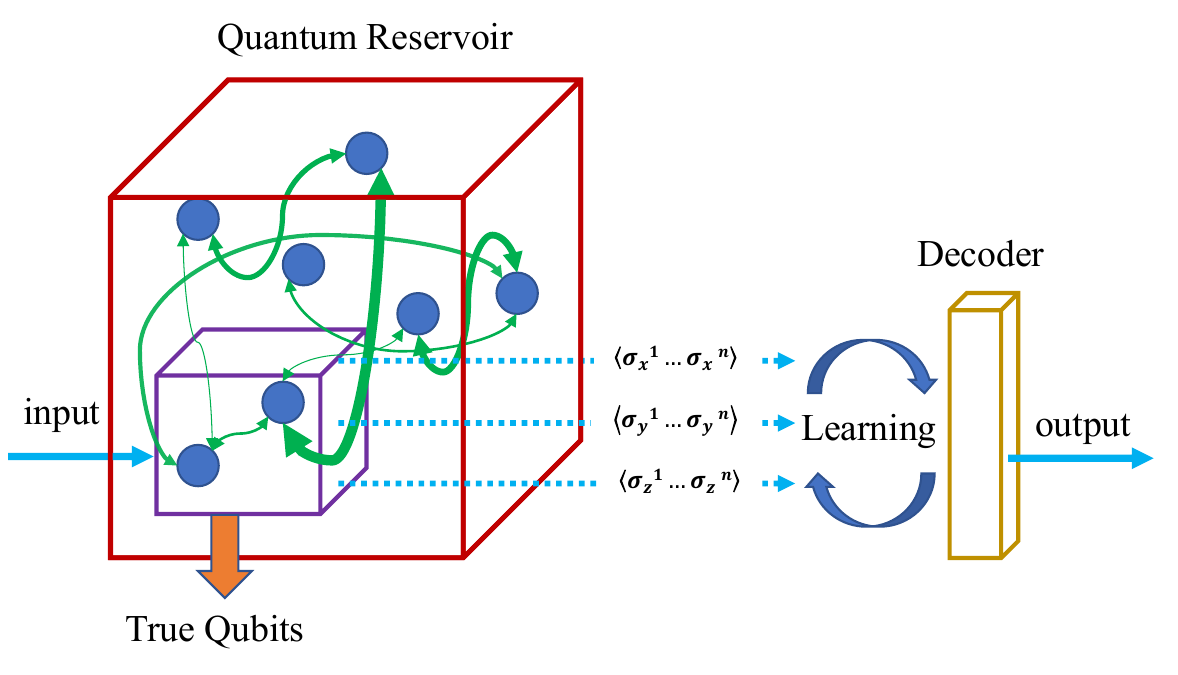}[h]
    \caption{\textbf{Schematic representation of the spin-based quantum reservoir computing.} The reservoir (red box) consists of N qubits (blue circles) which are randomly connected, and stay constant. At each step, the input data is fed to only a few of the qubits, true qubits (purple box). After data feeding, the correlation of Pauli matrices, $x,y,z$ from n qubits will be measured and sent to a classical decoder. For the next iteration, the true qubits will be traced out from the density matrix. The reservoir and the decoder are linked with weights that are the only ones that get modified during the learning process.}
\label{schematic}     
\end{figure}
\textit{Quantum} reservoir computing (QRC) has been on the rise in recent years, introducing to RC such counter-intuitive features of quantum physics as superposition and entanglement \cite{fujii_2017,fujii_2021}. Two distinct QRC models have been proposed: one is grounded in networks of qubits \cite{fujii_2021,Luchnikov_2019,Martnez_2020} and another in oscillator-based \cite{Govia_2021,Nokkala_2021,Dudas_2023} reservoirs. In this work we focus on the former model. Within this type of QR, transitions of basis states for quantum bits (qubits) are propelled by an input stream (Fig. \ref{schematic}), evolving over time through the reservoir's quantum dynamics \cite{Nakajima2019,martinez_2021}. Input data is fed into the \textit{true} nodes, typically in the form of a time series. For the readout, besides these true nodes, hidden nodes can also be used, which influence the time evolution of the true nodes. \\ \indent
The primary obstacle encountered by quantum computing and quantum machine learning is the noise present in quantum devices. Significant efforts have been dedicated to correcting or mitigating the resulting errors. Recent studies show that the presence of noise can improve the convergence of variational quantum algorithms \cite{Fry2023}. Further, it is suggested that quantum noise, under specific conditions, can enhance the efficacy of quantum reservoir computing \cite{suzuki2022,Domingo2023}.\\ \indent
For time-series data, the frequency and the stochasticity of the data play crucial roles in the performance of machine learning models. Time-series data, characterized by its sequential order, can vary in frequency from high-resolution milliseconds to monthly or yearly observations \cite{Abreu_2020,Ballarin_2023,Tanaka_2019}.  On the other hand, studies suggest potential advantages of quantum computing in systems characterized by high levels of stochasticity and randomness, though these advantages are not always measured against the most optimized classical solutions \cite{Dale2015,Blank2021,Korzekwa2021}. Experimental evidence tentatively indicates that quantum approaches to simulating stochastic processes might require less memory than traditional classical methods, under certain conditions \cite{Palsson2017}. Furthermore, preliminary findings propose that, by negotiating the trade-off between accuracy and memory usage, quantum models could potentially achieve comparable levels of accuracy with reduced memory requirements, or conversely, improve accuracy without increasing the memory footprint \cite{banchi}. However, these observations are context-dependent and should be considered with caution, as the comparisons are not universally applicable across all scenarios. \\ \indent
Beyond the inherent noise challenges, the salient aspects of quantumness —superposition and entanglement — are pivotal in understanding quantum systems. Recently, considerable research effort has been devoted to the role played by quantumness in giving rise to advantage in machine learning, whether in distributed learning over quantum networks \cite{gilboa_2023}, spin-network QRC \cite{gotting_2023}, or oscillator-based QRC \cite{motamedi_2023}. However, there remains much to be understood as to the physical circumstances conducing to such a quantum advantage. \\ \indent
Here, we conduct a thorough investigation of how the quantumness of Ising spin-networks, which we quantify by means of the logarithmic negativity measure of entanglement, relates to performance in QRC memory tasks; to wit, by investigating the ability
of the reservoir to reconstruct functions of past input, and the possible advantage of quantum entanglement for
this sort of task. We explore  how that relationship is affected by dissipation, of which some amount has been observed to enhance performance \cite{Domingo2023, suzuki_2022, gotting_2023}; as well as by the frequency of the input signal, which we show here has great implications for the computational advantage granted by entanglement. We find that the presence of that advantage is frequency-dependent in the presence of dissipation but is always present in the unitary reservoir. We attribute this phenomenon to the new timescale introduced by dissipation, which we conclude must be longer than the timescale over which the input varies in order for the entanglement advantage to be realized. In other words, quantum memory must survive for long enough in the system for the input to manifest its temporal features, which is determined by the frequency scale of the input signal. We also find quantum entanglement to aid in the remembrance of more features. In the high-frequency cases where this happens, we observe diminishing returns on the entanglement advantage; the positive impact of entanglement on performance exhibits saturation behavior.  \\ \indent
The remainder of this paper is organized as follows: in section \ref{meth} we describe our dynamical models and methodology for data extraction and coarse-graining. We present and discuss our results in section \ref{res}, and finally outline our conclusions in section \ref{conc}.

\section{Model and Methodology}\label{meth}
\subsection{Physical System}\label{sys}
The quantum reservoir we employ is a network of  $N=4$ qubits represented by the transverse-field Ising model \cite{stinchcombe_1973,pfeuty_1971}, of which the Hamiltonian reads
\begin{eqnarray}\label{ham}
\hat H = \sum_{i>j=1}^{N} J_{ij} \hat\sigma_{i}^{x} \hat\sigma_{j}^{x} + h \sum_{i=1}^{N} \hat\sigma_{i}^{z},
\end{eqnarray}
where $\hat\sigma_i^a$ ($a=x,y,z$) are the Pauli operators, $h$ is the transverse magnetic field, and $J_{ij}$ are randomly chosen from a uniform distribution in the interval $[-J_s/2,J_s/2]$.
We consider an open system in general, of which the Markovian dynamics are given by the Lindblad master equation
\begin{align}\label{mastereq}
\frac{d \hat\rho}{dt} = \mathcal{\hat L} \hat\rho = -i[\hat H, \hat\rho] + \Gamma \sum_{i=1}^{N} \left( \hat L_i \hat \rho \hat L_i^\dagger - \frac{1}{2} \{\hat L_i^\dagger \hat L_i, \hat \rho\} \right),
\end{align}
where $\Gamma$ controls the strength of the dissipation of our high-temperature decoherence channel \cite{breuer_2002} with $\{\hat L_i\}$ being the raising and lowering operators of the system given by
\begin{eqnarray}\label{sig}
\hat \sigma^+ = \frac{1}{2}(\hat\sigma^x + i\hat\sigma^y), \\
\hat\sigma^- = \frac{1}{2}(\hat\sigma^x - i\hat\sigma^y).
\end{eqnarray}
The unitary system undergoes a dynamical phase transition between an ergodic phase and a many-body-localized phase when the ratio $h/J_s$ falls within a certain range, as reported in the phase diagram in Ref. \cite{martinez_2021}. 
\begin{figure}[h]
\centering
\includegraphics[width=0.5\textwidth]{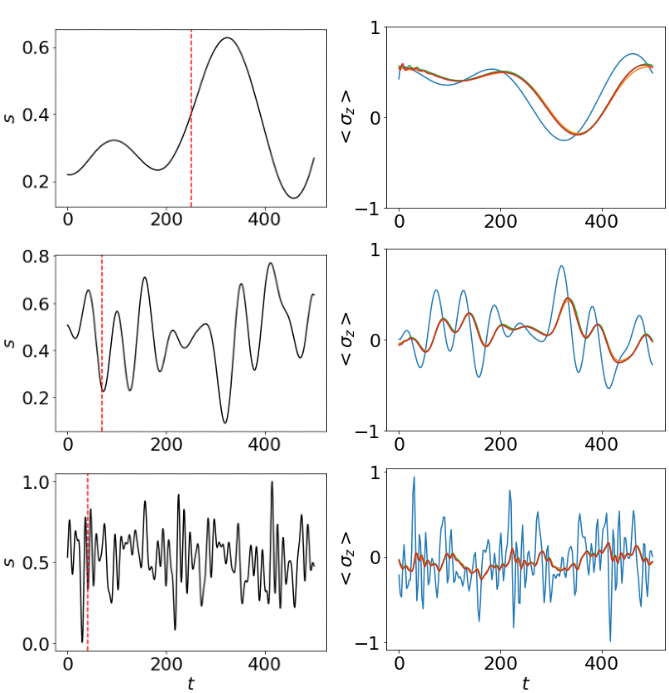}
    \caption{(left) Examples of input sequences at three different frequency scales $f=0.2$ (top), $f=1$ (middle), $f=5$ (bottom). The red vertical dashed line corresponds to the maximum time delay at which the reservoir is successful in remembering past input, based on the results we report later in the manuscript. (right) The corresponding outputs of the reservoir at an interaction strength of $J_s=1$, a transverse field of $h=2$, an injection period of $\Delta=2.5$, and a dissipation strength of $\Gamma=0.01$.}
\label{input}     
\end{figure}
\subsection{Input and Training}\label{inp}
An input signal $s_k$ with a frequency scale $f$ is generated as follows. 20 frequencies $\left\{f_i\right\}$ are chosen with equal linear spacing in the interval $\left[ f/5000, f/50 \right]$, thereby producing the series 
\begin{align}\label{sk}
s_k = \sum_i^{20} \sin\left(2\pi f_i  t_k + 2 \pi \zeta\right),
\end{align}
where $t_k$ is the time after $k$ time steps, and $\zeta$ is a random number uniformly chosen in $\left[0,1\right]$. All input functions are rescaled to have a range between 0 and 1. Examples of such series at different frequency scales $f$ are shown in the left panels of Fig. \ref{input}, while the right panels show the corresponding outputs for the dissipative reservoir at a representative choice of parameters. \\ \indent 
The input is injected into the systems by means of reinitializing the state of one qubit that we call qubit 1, a widely utilized form of input encoding \cite{mujal_2021}: we trace out the qubit in question and prepare it in a state $|\psi_{s_k}\rangle = \sqrt{1 - s_k} |0\rangle + \sqrt{s_k} |1\rangle$, such that
\begin{align}\label{fourier2}
\rho \rightarrow |\psi_{s_k}\rangle \langle\psi_{s_k}| \otimes \text{Tr}_{1}[\rho]
\end{align}
every $\Delta t =L\delta t$, where $L$ is the number of time steps between input injections. We choose $\Delta t$ such that $h\Delta t$ lies in the range prescribed in Ref. \cite{martinez_2020} for rich linear and non-linear dynamical behaviour. \\ \indent
We employ time multiplexing, where outputs are read at time intervals of $\Delta t/V$, thus allowing for $NV$ virtual nodes. The signals extracted from these nodes, here restricted to $\langle\sigma_z\rangle$, are trained in a linear regression to produce the desired function of the input, either the fundamental linear memory task \cite{carroll_2022} $\bar{y}_k = s_{k-\tau}$ ($\bar{y}_k$ being the target), or a non-linear NARMA-n task, which is described and discussed in the Appendix. \\ \indent
The system is trained on multiple sequences and tested using a different sequence, $f$ being the same between the training and testing phases. Tikhonov regularization is employed with a regularization parameter that is allowed to slide to maximize performance. The performance of the reservoir in a memory task with time delay $\tau$, here referred to as the memory capacity, is quantified as
\begin{align}\label{memcap}
C^{\tau}_{STM} = \frac{\text{cov}^2(y, \bar{y}^\tau)}{\sigma_{y}^2 \sigma_{\bar{y}^\tau}^2},
\end{align}
where  $y$ is the output signal of the reservoir and $\bar{y}^\tau$ is the target function at a time delay $\tau$. The memory capacity is subject to the fading memory property of the reservoir computer \cite{dambre_2012}; it must vanish for sufficiently large $\tau$. Thus the total memory capacity of the reservoir may be defined as
\begin{align}\label{totmemcap}
C_{\text{STM}} = \sum_{\tau=0}^{\infty} C_{\text{STM}}^{\tau}.
\end{align}
\subsection{Entanglement}\label{pca}

\begin{figure}[h]
\centering
\includegraphics[width=0.3\textwidth]{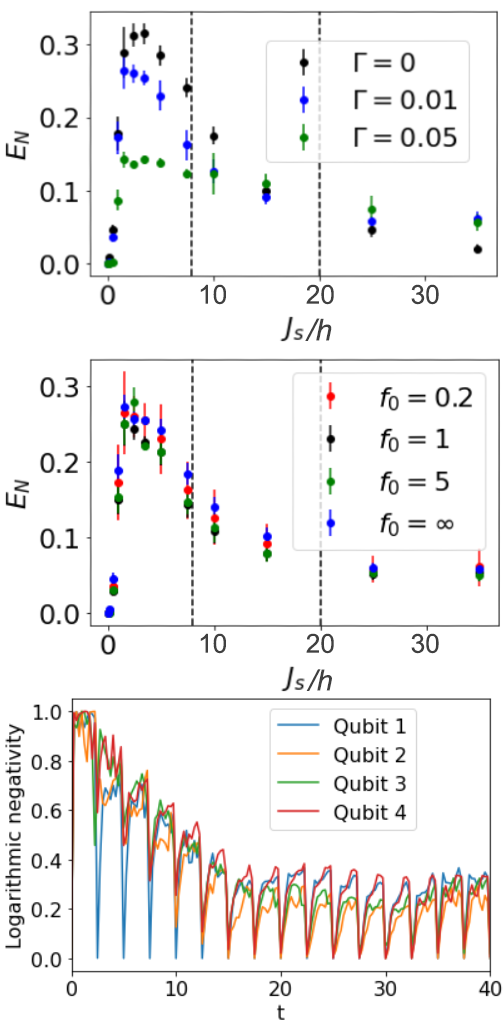}
    \caption{(top) Entanglement vs. interaction strength for the reservoir at a transverse field of $h=2$, and an injection period of $\Delta t=2.5$.  The input of frequency of the signal is fixed at $f=0.2$ while the dissipation strength is varied. (middle) Same as top but the dissipation strength is fixed at $\Gamma=0.01$ and the input frequency is varied. The black lines mark the boundary of the dynamical phase transition reported in \cite{martinez_2021}. (bottom) An example of the logarithmic negativity as a function of time - the entanglement abruptly drops whenever the input is injected. Each color corresponds to the bipartition in which the respective qubit labeled in the legend is isolated. }
\label{ent}     
\end{figure}

 Quantum entanglement is measured by means of the logarithmic negativity \cite{vidal_2002,plenio_2005}
\begin{equation}\label{logent}
E_N(\rho) = \log_2 \|\rho^{\Gamma_A}\|_1,
\end{equation}
where $\| \cdot \|_1$ denotes the trace norm, $\rho^{\Gamma_A}$ is the partial transpose with respect to subsystem A; all bipartitions are averaged over. An example of the behaviour of this quantity is shown in Fig. \ref{ent} (top inset), where 4 different bipartitions are shown, each isolating one of the 4 qubits. The entanglement experiences a sudden drop each time the input enters the system, especially the bipartition isolating qubit 1, into which the input is injected.

\section{Results and Discussion}\label{res}
\begin{figure}[t]
\centering
\includegraphics[width=0.47\textwidth]{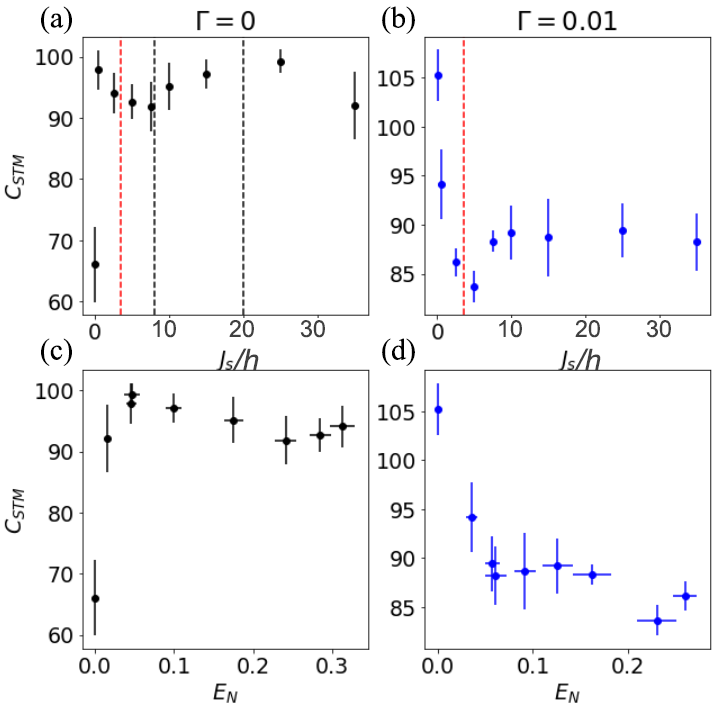}
    \caption{(top) Total memory capacity in the linear memory task vs. interaction strength for the unitary (a) reservoir compared with the dissipative reservoir $\Gamma=0.01$ (b), at a transverse field of $h=2$, an injection period of $\Delta t=2.5$, and the input of frequency of the signal is fixed at $f=0.2$. The black lines mark the boundary of the dynamical phase transition of the unitary reservoir. The red line corresponds to the location of maximum entanglement. (bottom) The same but vs. entanglement.}
\label{f0p2lin}     
\end{figure}

We commence by presenting the results for the logarithmic negativity measure of entanglement, which we employ to quantify quantumness. Those results are plotted in Fig. \ref{ent} against the interaction strength for a variety of dissipation strengths and input of frequencies, at a transverse field of $h=2$, and an injection period of $\Delta t=2.5$. In this figure and all subsequent figures, the critical region between the black lines is that in which the dynamical phase transition takes place in the unitary reservoir between thermalization, to the left of the region, and localization, to the right \cite{martinez_2021}. It should be noted here that the the lowest interaction strengths in all subsequent plots are small but finite. \\ \indent 
A few interesting properties stand out. Firstly, we note that the logarithmic negativity is at its highest to the left of the dynamical phase transition, well into the ergodic regime, where thermalization leads to the spreading of entanglement throughout the system. On the other hand, entanglement is lowest at the extremes: at very high interaction strength the system is deep into the many-body localized phase and the spread of entanglement is stifled, and at very low interaction strength the spins interact too weakly to be entangled. The interaction strength that maximizes entanglement is marked with a red line in all subsequent figures. We also note that the average entanglement goes down as we ramp up the dissipation, albeit with a qualitatively similar shape as a function of interaction strength; it peaks at virtually the same location and dies at the extremes. However, the entanglement appears to be largely insensitive to the frequency of the input signal. \\ \indent
Now, in order to set the stage, we go to the special case of the unitary reservoir at low frequency, of which Fig. \ref{f0p2lin}a presents some intriguing observations. There, the total memory capacity is computed for the linear memory task. The results for the the non-linear NARMA task, qualitatively similar, are presented in Fig. \ref{f0p2narm} in the appendix. Fig. \ref{f0p2lin}a shows the evolution of the performance as we ramp up the interaction strength $J_s$: deep in the ergodic regime (low $J_s$) the performance is poor, but improves rapidly as the system approaches the region of the dynamical phase transition indicated by the back lines. \\ \indent
It's certainly clear from the left panels of Fig. \ref{f0p2lin} that no entanglement leads to poor performance - but too much entanglement doesn't seem to be optimal either, for the maximum of entanglement appears to lead to rather a dip in performance. This idea is emphasized in Fig. \ref{f0p2lin}c, in which the memory capacity is plotted this time against entanglement. Performance is undoubtedly enhanced as entanglement rises from zero - consistent with the presence of an entanglement advantage - but this increase does not continue when the entanglement becomes too great. This suggests that in the case of the unitary reservoir at low frequency, there exists an entanglement advantage with diminishing returns. While an interesting result, it should be noted that the behavior of the unitary reservoir at low frequency is merely a special case to lay the groundwork for examining the more general case of a reservoir with finite dissipation subjected to an input of arbitrary frequency. \\ \indent
The behaviour of the non-unitary reservoir is investigated by switching on the dissipation $\Gamma$ in eq. (\ref{mastereq}). We start at the original conditions of $h=2$, $\Delta t=2.5$, and the low frequency of $f=0.2$. As shown in the top panel of Fig. \ref{ent}, the ramping of dissipation lowers the entanglement curves. However, the entanglement is clearly insensitive to the input frequency at a given dissipation strength, as is shown in Fig. \ref{ent} (bottom).  Here, it's important to remember that the phase boundaries were computed in Ref. \cite{martinez_2021} for the \textit{unitary} reservoir. \\ \indent
\begin{figure}[t]
\centering
\includegraphics[width=0.45\textwidth]{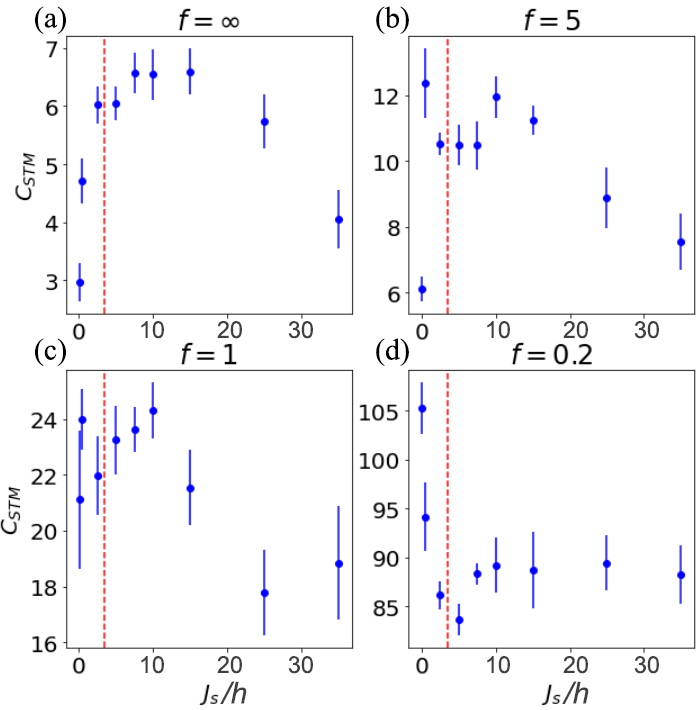}
    \caption{Total memory capacity vs. interaction strength for the dissipative ($\Gamma=0.01$) reservoir at a transverse field of $h=2$, and an injection period of $\Delta t=2.5$, subjected to a variety of input of frequencies $f$, in linear memory tasks at time delay $\tau$. The red line corresponds to the location of maximum entanglement.}
\label{linmemjg0p01}     
\end{figure}
\begin{figure}[h]
\centering
\includegraphics[width=0.47\textwidth]{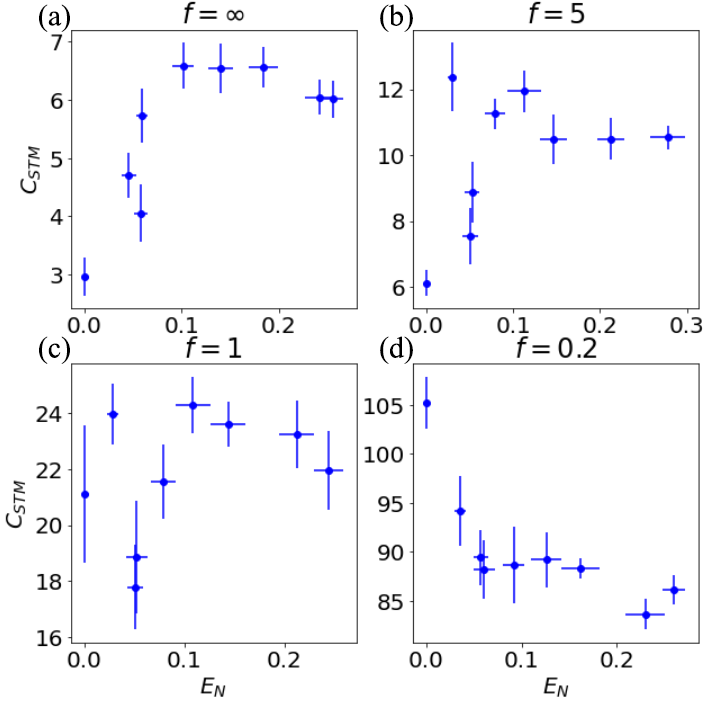}
    \caption{Total memory capacity vs. entanglement for the dissipative ($\Gamma=0.01$) reservoir at a transverse field of $h=2$, and an injection period of $\Delta t=2.5$, subjected to a variety of input of frequencies $f$, in linear memory tasks. }
\label{linmementg0p01}     
\end{figure}
\begin{figure}[h]
\centering
\includegraphics[width=0.47\textwidth]{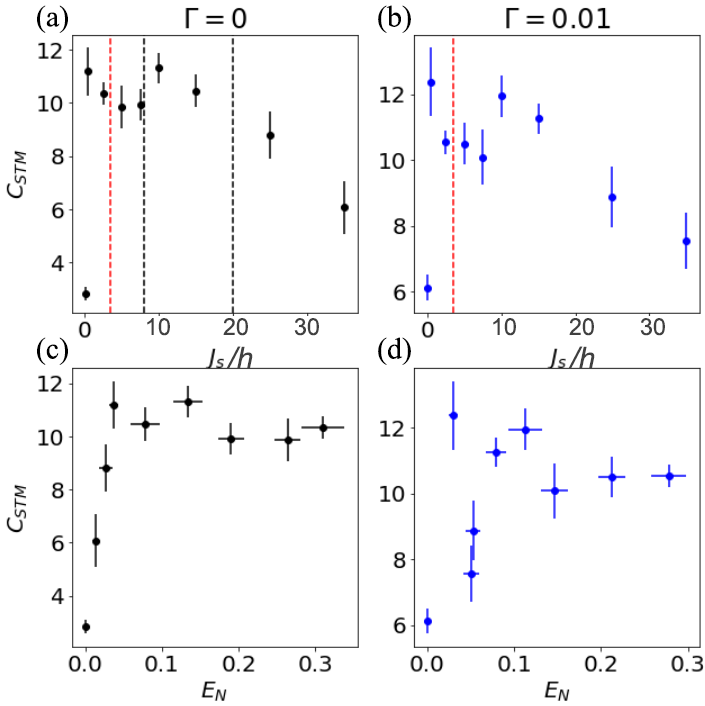}
    \caption{(top) Total memory capacity in the linear memory task vs. interaction strength for the reservoir at a transverse field of $h=2$, and an injection period of $\Delta t=2.5$, subjected to an input frequency of $f=5$, at the dissipation strengths $\Gamma=0$ (a) and $\Gamma=0.01$ (b). The black lines mark the boundary of the dynamical phase transition of the {\em unitary} reservoir. The red line corresponds to the location of maximum entanglement. (bottom) The same but vs. entanglement.}
\label{memf5}     
\end{figure}
\begin{figure}[h]
\centering
\includegraphics[width=0.5\textwidth]{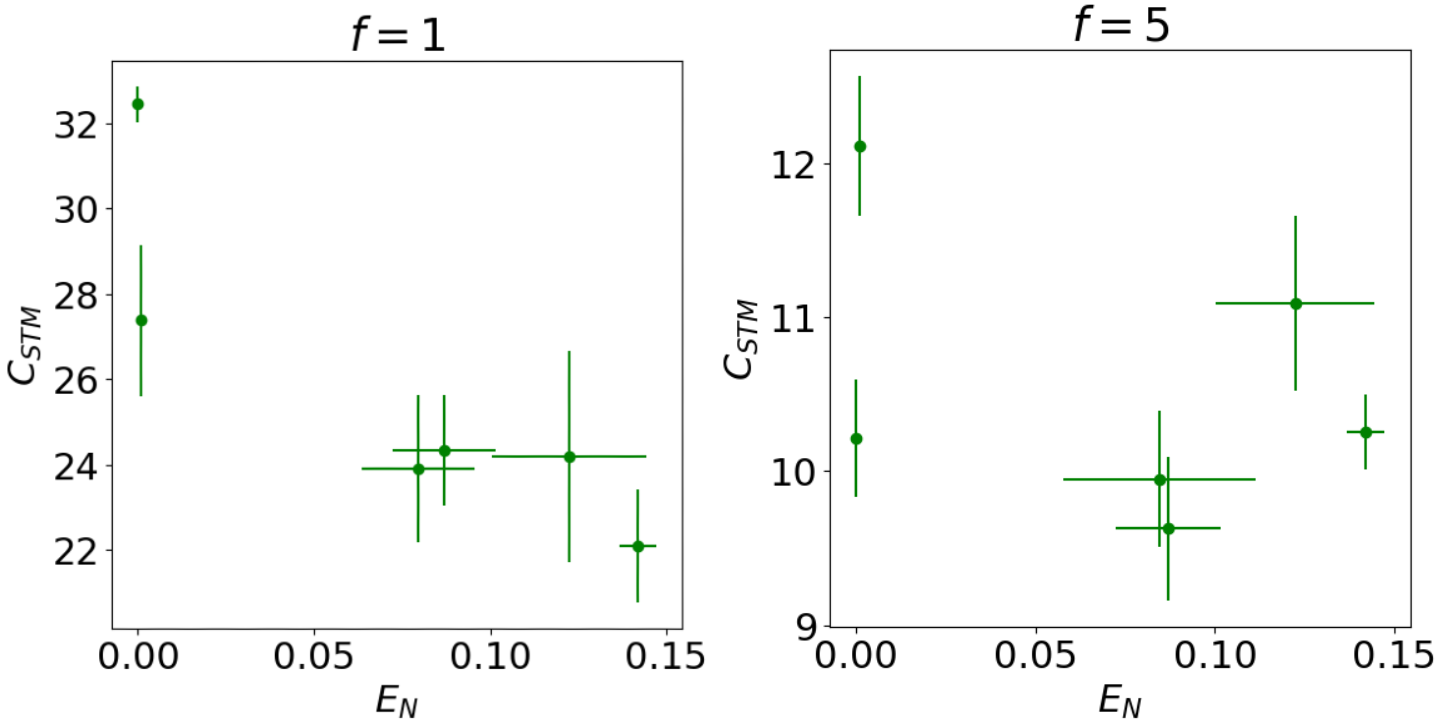}
    \caption{Total memory capacity in the linear memory task vs. entanglement for the dissipative reservoir $\Gamma=0.05$ at a transverse field of $h=2$ and an injection period of $\Delta t=2.5$, and an input signal of frequencies $f=1$ (left) and $f=5$ (right).}
\label{g0p05}     
\end{figure}
\begin{figure}[h]
\centering
\includegraphics[width=0.47\textwidth]{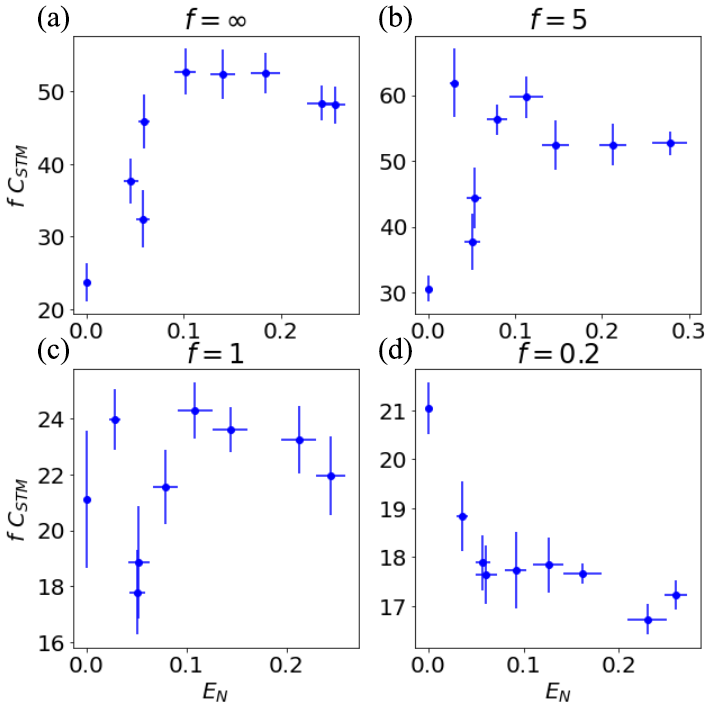}
    \caption{The same as Fig. \ref{linmementg0p01} but with the y-axis of the frequency rescaled by a factor of the input frequency.}
\label{fc}     
\end{figure}
The behaviour of the memory capacity in the dissipative reservoir subject to an input frequency of $f=0.2$, as illustrated with an example in Fig. \ref{f0p2lin}b, is rather odd: it develops an abrupt and steep rise at very low interaction strength, where there is little entanglement. Such a jump in behaviour, present in both the linear memory task shown here, and the NARMA task in Fig. \ref{f0p2narm} in the appendix, makes for a rather striking sight upon comparing the black curves, with zero dissipation, to the blue curves, where even a small amount of dissipation is introduced. It indicates a discontinuity in the dissipative reservoir at low interaction strength and calls into question the presence of the entanglement advantage that is very clear in the case of the unitary reservoir, as illustrated in Fig. \ref{f0p2lin}d, where the total memory capacity may be seen to decline as the entanglement rises.\\ \indent
The key to unravelling this quandary lies in generalizing the investigation even further to study the effect of the input frequency. This is accomplished in Fig. \ref{linmemjg0p01}, which reports the linear memory capacities at the usual conditions of field and injection period, and at a dissipation of $\Gamma=0.01$, but computed for different input frequencies (the corresponding NARMA results are presented in Fig. \ref{narmajg0p01}). Here, $f=\infty$ corresponds to a sequence of random floats between 0 and 1 with no continuity. The entanglement advantage is decidedly restored by raising the input frequency, as the performance starts exhibiting a peak at intermediate connectivities. \\ \indent
The return of the entanglement advantage is corroborated by Fig. \ref{linmementg0p01} (whose NARMA counterpart is Fig. \ref{narmaentg0p01}) , where the total memory capacity as a function of entanglement may be seen to evolve from one extreme to the other. At low frequency, there is a very clear entanglement disadvantage as previously observed, as maximum memory performance occurs where entanglement is smallest and quantum effects are suppressed. As the frequency of the input signal rises, however, the performance enhancement at low entanglement gradually fades. In those high-frequency cases, the entangled systems are significantly better than the non-entangled systems, albeit with diminishing returns on the benefit of entanglement, since performance saturates as entanglement rises. This suggests that there may be an optimal amount of entanglement for memory performance.\\ \indent
One interpretation of this phenomenon is that, while as low a frequency as $0.2$ was perfectly fine for the unitary reservoir, in the presence of a finite amount of dissipation it is so low as to give rise to an incongruence of covariance dimension and performance, as may be seen by comparing Fig. \ref{f0p2lin} and Fig. \ref{dim}, and a fading of the entanglement advantage. This reveals that the presence of the entanglement advantage is dependent on the input frequency - however, this scarcely detracts from the significance of the entanglement advantage, as a higher frequency signal is more difficult to process by virtue of its having more features per unit time. This is consistent with the observation of quantum advantages in highly stochastic and random systems \cite{Dale2015,Blank2021,Korzekwa2021}. \\ \indent
The revival of the entanglement advantage in the dissipative reservoir at high frequency raises an important question: how does the unitary reservoir operate at high frequency? Fig. \ref{memf5} offers an answer to that question by showing the total memory capacities in the linear memory task at $f=5$ (NARMA, once again qualitatively similar, is presented in Fig. \ref{f5narm} in the appendix), comparing the unitary reservoir and the case with a small amount of dissipation ($\Gamma=0.01$). The entanglement advantage is very much present in both cases, as shown in the bottom panels of Fig. \ref{memf5}. Also noteworthy is that the small amount of dissipation seems to result in a small improvement in memory capacity especially at low connectivity (obvious on comparing the leftmost points in the top panels of Fig. \ref{memf5}), which is consistent with the observation reported in \cite{gotting_2023}, and the ideas presented in \cite{Domingo2023,suzuki_2022}. \\ \indent
The frequency-dependent nature of the entanglement advantage in the dissipative case, as compared to the omnipresence of the entanglement advantage in the unitary case, may be attributed to the arising of a new timescale upon switching on dissipation. We introduce dissipation into the system by allowing $\Gamma$ in eq. \ref{mastereq} to acquire a non-zero value, which gives rise to a new timescale in the system: $1/\Gamma$. Our observations indicate that input signals with a timescale longer than $1/\Gamma$ do no benefit from entanglement. This may be seen in the top panels of Fig. \ref{input}, where it's clear that the lowest frequency input signal varies over a timescale of roughly $200$, which is longer than the dissipative timescale $1/\Gamma=1/0.01=100$. All higher frequency inputs, however, vary over a shorter time scale than $1/\Gamma$, and thus derive benefit from entanglement. However, in the unitary case, the dissipative timescale is of course infinite, and thus all input signals - even the lowest frequency one - vary over a shorter timescale and the entanglement advantage is always present. We now to turn to higher dissipation in order to verify this idea that the timescale of the input signals for which there is an entanglement advantage is dictated by the strength of dissipation $\Gamma$. We thus extend our analysis by raising the value of $\Gamma$ from 0.01 to 0.05, and explore the corresponding performance curves as a function of entanglement in Fig. \ref{g0p05}. Indeed, these results confirm that upon lowering the timescale of dissipation by raising $\Gamma$ to 0.05, an input signal of $f=1$ becomes slow enough that an entanglement {\it disadvantage} manifests itself. This stands in contrast to the lower $\Gamma$ in Fig. \ref{linmementg0p01}, where $f=1$ is fast enough to be a transition point between an advantage and a disadvantage. This is also apparent by comparing the $f=5$ panels in both figures; $f=5$ is no longer as clear a case of entanglement advantage when the dissipation is raised, but appears to be in the regime of entanglement indifference. These observations indicate that quantum memory in the system, which decays on a timescale of $1/\Gamma$, only confers a memory advantage in the presence of inputs whose features manifest themselves before that quantum memory decays. For inputs with longer timescales, classical memory takes over and the entanglement advantage wanes. This constitutes a condition that must be satisfied before the entanglement advantage may be realized. \\ \indent
Finally, we recast the data in Fig. \ref{linmementg0p01} in a new light, namely by multiplying the total memory capacities by the input frequency, as shown in Fig. \ref{fc}. The point of this recasting is to render the comparison of the memory capacities between different frequencies more fair; lower frequency tasks are much easier because they contain fewer temporal features, and so this rescaling gives an estimate of the memory performance with respect to the number of features per unit time. In the case of $f_\infty$, which again is merely shorthand for the sequence of random floats, we rescaled the total memory capacity by comparing the frequency of stationary points in that case with the frequency of stationary points at $f=5$. The additional insight revealed Fig. \ref{fc} is that as we raise the frequency from $f=0.2$ (where the best performance is by the non-entangled system), the points corresponding to the entangled systems are the ones that rise up to effect the entanglement advantage, rather than the non-entangled system falling off. Indeed, the performance of the unentangled system remains quite consistent throughout the range of input frequencies. This suggests that quantum entanglement empowers the system to remember more temporal features.
\section{Conclusions and Outlook}\label{conc}
We studied the performance of a spin network QRC system in linear and non-linear memory tasks, and how it relates to such things as the frequency of the input signal, the presence of dissipation in the system, and quantumness as measured by entanglement. \\ \indent
In the open system, the most general case, it's clear that the presence of the entanglement advantage is frequency-dependent in a manner which is decreed by the strength of dissipation in the system. A task with too low a frequency of input signal is too slow to derive any benefit from quantum entanglement entanglement, and the classical limit with weak interactions and no entanglement performs better in linear and non-linear memory tasks. On the other hand, in the presence of a sufficiently involved signal which fluctuates over timescales that are shorter than the timescale introduced by the strength of dissipation, quantum entanglement confers a clear advantage upon the system in both types of tasks. Put differently, the presence of the quantum entanglement advantage is contingent on the quantum memory's survival for long enough that the input may have time to exhibit its temporal features - the number of those temporal features is greater for higher-frequency input signals. The more features a signal possesses over the timescale for which quantum memory persists, the more benefit conferred by quantum entanglement. In that sense, we conclude that quantum entanglement enhances the ability of the spin-network quantum reservoir to remember more temporal features.
\\ \indent
Possible extensions of this study include the integration of neuromorphic elements into the reservoir. One such possibility involves exploring different reservoir architectures inspired by brain networks, which have been shown to be amenable to being coarse-grained to appropriately large scales while retaining much of their dynamical behavior \cite{kora_2023cg}.  Another possibility involves studying the potential dependence of the performance of the system on system size, i.e., how it scales with the number of qubits. Furthermore, there is the prospect of investigating spin network reservoirs handling tasks that can learn functions of multiple inputs, in a manner inspired by the brain's reception of multiple sensory stimuli and processing them in conjunction.   \\ \indent
One among several promising avenues is that of quantum computation and simulation with neutral atoms, with which the possibilities are myriad \cite{henriet_2020}; Rydberg atoms are exceptional candidates for the implementation of QRC \cite{araiza_2022} on account of their strong interactions, controllable states, and long coherence times – characteristics that make them ideal for creating highly interactive and adaptable quantum reservoirs. Neutral atoms may be used to implement both the quantum Ising model \cite{scholl_2021}, which sits at the foundation of spin-network QRC; and the XY model \cite{barredo_2015, pineiro_2018}, which would be a fascinating platform on which to implement an additional level of complexity and non-linearity in a quantum reservoir. This is one among several platforms that are well-suited for QRC implementation, such as superconducting qubits \cite{suzuki_2022}, photonics \cite{Garc_2023}, and trapped ions \cite{haffner_2008}. Making contact with such implementations will be the topic of future investigation.
\section{Acknowledgments}\label{ack}
This work was supported by the National Research Council through its Applied Quantum Computing Challenge Program, the Natural Sciences and Engineering Research Council (NSERC) of Canada through its NSERC Discovery Grant Program, the Alberta Major Innovation Fund, and Quantum City. We would also like to thank Aaron Goldberg for the useful discussions and feedback. \\

\bibliography{biblio}

\section{Appendix}
\subsection{Principal Component Analysis} 
\begin{figure}[h]
\centering
\includegraphics[width=0.4\textwidth]{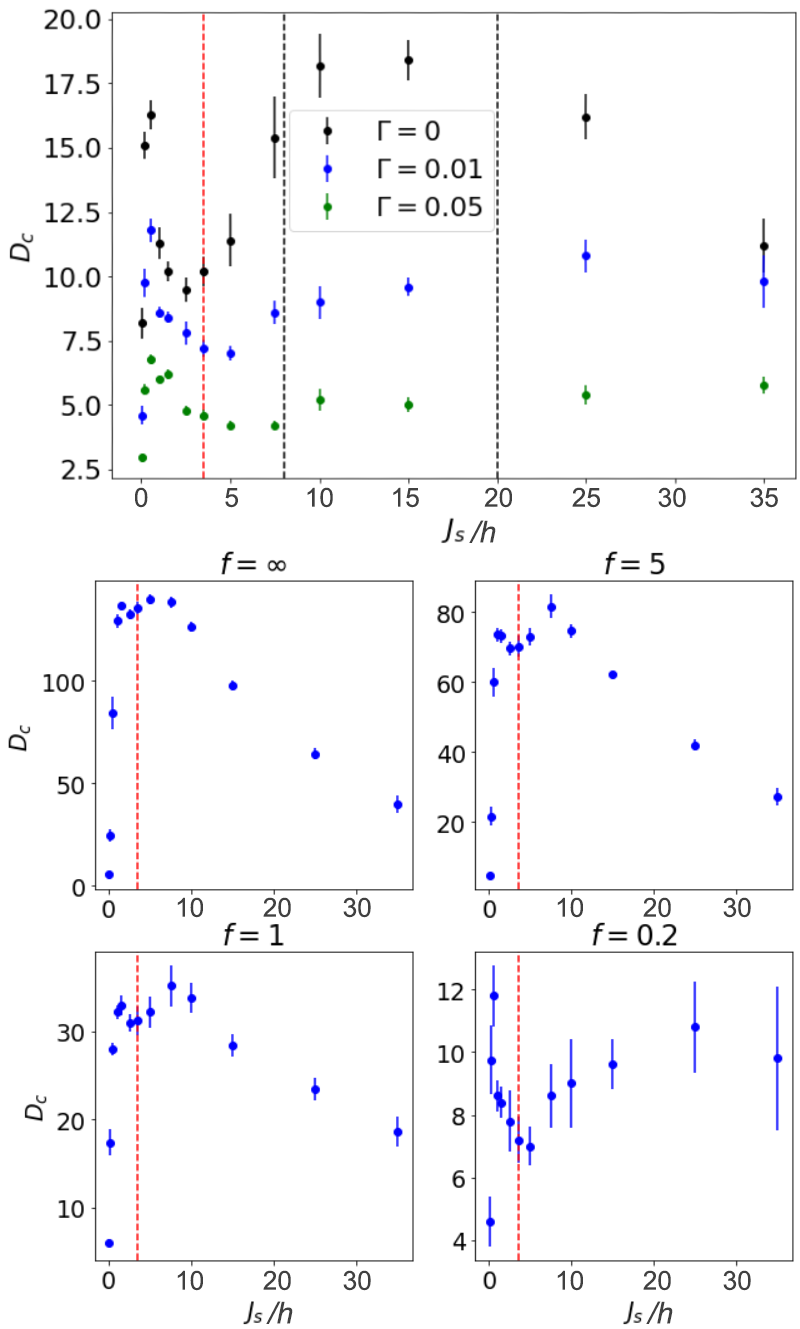}
    \caption{Covariance dimension vs. interaction strength at the same sets of conditions as Fig. \ref{ent}. The black lines mark the boundary of the dynamical phase transition reported in \cite{martinez_2021}. The red line corresponds to the location of maximum entanglement.}
\label{dim}     
\end{figure}
To estimate the dimensionality of the manifold to which the dynamics is restricted, principal component analysis is performed as in \cite{gotting_2023}. Let the quantum dynamics of the system be represented by the trajectory of the vector evolution $\bf X = (x_0, x_1, \ldots)$, where $\bf x_i$ are the density matrices of the system sampled at the multiplexed intervals $\Delta t/V$. A random time index is chosen and a cluster of $d+1$ nearest neighbours is determined and combined into a matrix $X_{i0}$, of which the covariance matrix is computed
\begin{equation}\label{cov}
C_{\bf{X}_{i_0}} = \frac{1}{N_d - 1} (\bf{X}_{i_0} - \bar{\bf{X}}_{i_0})(\bf{X}_{i_0} - \bar{\bf{X}}_{i_0})^T
\end{equation}
The covariance dimension is the number of principal components of $C_{X_{i_0}}$ greater than $\epsilon=10^{-6}$ (which is fixed at a small value for all points), and is averaged over many iterations \cite{gotting_2023}. \\ \indent
\begin{figure}[h]
\centering
\includegraphics[width=0.4\textwidth]{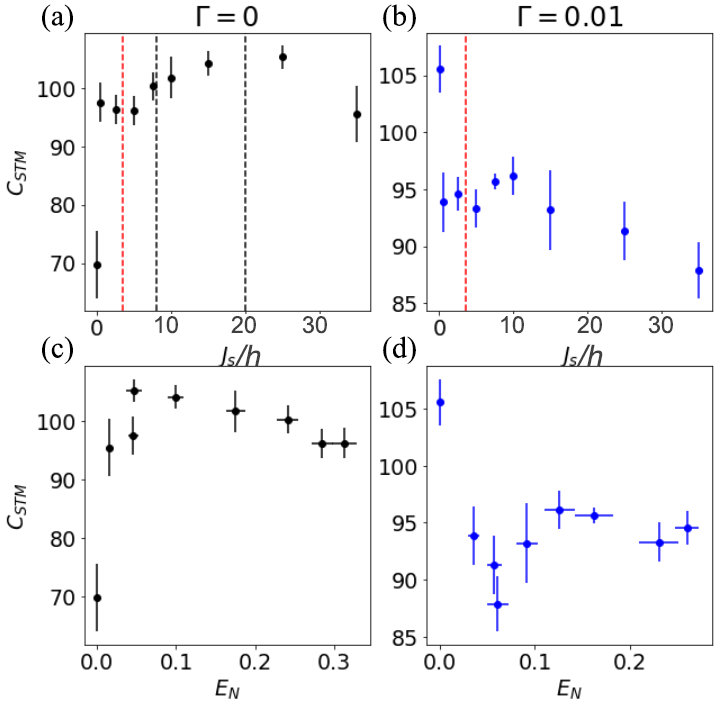}
    \caption{Same as Fig. \ref{f0p2lin} but for the NARMA task.}
\label{f0p2narm}     
\end{figure}

\begin{figure}[h]
\centering
\includegraphics[width=0.4\textwidth]{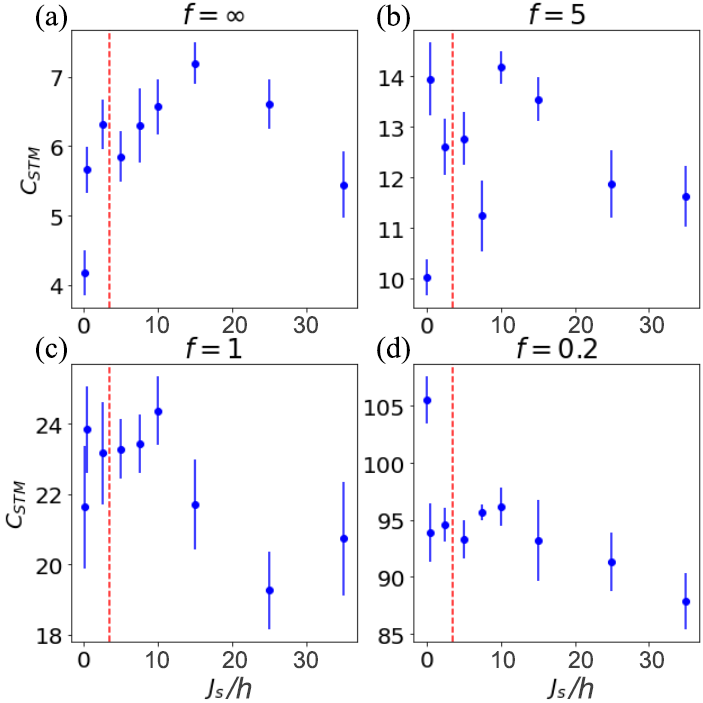}
    \caption{Same as Fig. \ref{linmemjg0p01} but for the NARMA task.}
\label{narmajg0p01}     
\end{figure}
\begin{figure}[h]
\centering
\includegraphics[width=0.4\textwidth]{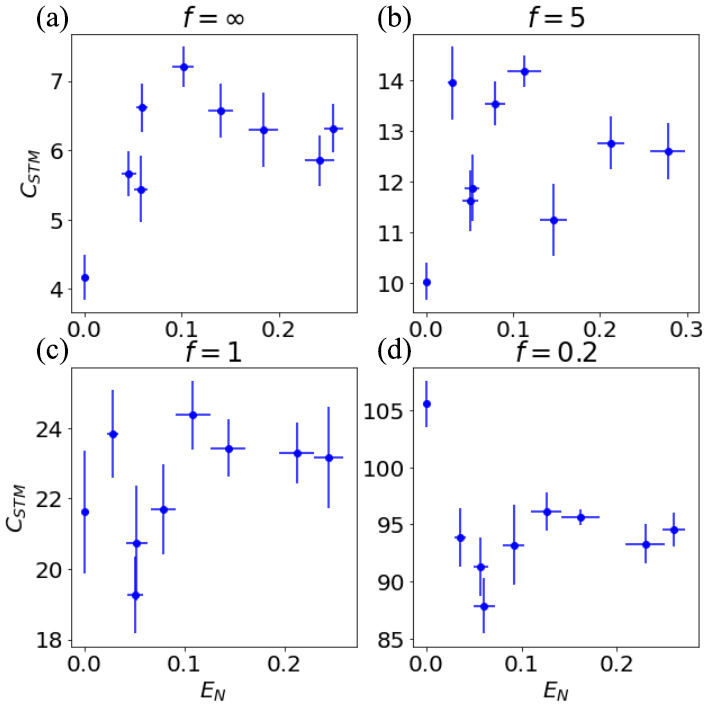}
    \caption{Same as Fig. \ref{linmemjg0p01} but for the NARMA task.}
\label{narmaentg0p01}     
\end{figure}
\begin{figure}[h]
\centering
\includegraphics[width=0.4\textwidth]{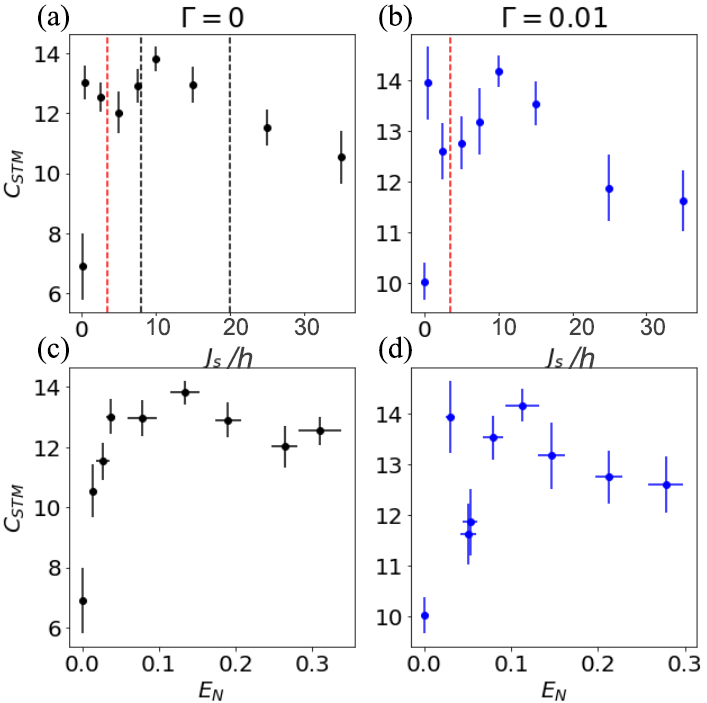}
    \caption{Same as Fig. \ref{memf5} but for the NARMA tasks}
\label{f5narm}     
\end{figure}
The relevant PCA results are summarized in Fig. \ref{dim}. The bottom panels ($f=0.2$) makes it clear that the small dip in performance observed in Fig. \ref{f0p2lin} is rather amplified, as the dimensionality experiences a much more pronounced dip at that region. There is also a substantial boost in the dimensionality of the unitary reservoir conferred by criticality, which is consistent with the performance boost at the dynamical phase transition observed in \cite{martinez_2021}. As dissipation is raised, covariance dimension is suppressed. \\ \indent
The top panel of Fig. \ref{dim} highlights the  ostensible strangeness of the results in Fig. \ref{f0p2lin}; upon comparing the two figures, it's clear that there is a congruence between covariance dimension and memory performance of the {\em unitary} reservoir, as they both rise from zero, peak, experience a dip, peak again near criticality, and then fall again. When dissipation is switched on, one can see the anomaly in Fig. \ref{f0p2lin}, where performance blows up at low interaction strength all of a sudden, which is not followed by a corresponding rise in dimensionality in the top panel of Fig. \ref{dim}. This may be understood, as explained in the main text, in terms of the new timescale introduced by the dissipation strength $\Gamma$.\\ \indent
The congruence of performance and dimensionality in the dissipative reservoir is re-established at sufficiently high frequency, when the entanglement advantage is restored. This may be seen by comparing Fig. \ref{dim} and  Fig \ref{linmemjg0p01}. The dimensionality curves, consistently familiar in shape, are only qualitatively resembled by the memory capacity curves at sufficiently high frequency. In short, our results indicate that the congruence between performance and dimensionality goes hand-in-hand with the presence of the entanglement advantage. Such an entanglement advantage, as we conclude in the main text, is only present when the input signal varies over a timescale which is shorter than the timescale of dissipation, $1/\Gamma$.\\ \indent
\subsection{The Non-Linear NARMA Task} 
The NARMA-n task \cite{atiya_2000} is given by 
\begin{align}\label{narma}
\bar{y}_{k+1} = \alpha \bar{y}_k + \beta \bar{y}_k \left( \sum_{j=1}^{n} \bar{y}_{k-j} \right) + \gamma s_{k-n} s_{k-1} + \delta,
\end{align}
where $n$ is the maximum time delay and $\alpha=0.3$, $\beta=0.002$, $\gamma=1.5$, $\delta=0.1$, convergent at a great range of frequencies. Figs. \ref{f0p2narm}, \ref{narmajg0p01}, \ref{narmaentg0p01}, \ref{f5narm} present the NARMA counterparts of the results of the main text, which are qualitatively similar in nature.

\end{document}